\input harvmac\skip0=\baselineskip
\def\p{\partial}


\lref\orv{
A.~Okounkov, N.~Reshetikhin and C.~Vafa,
``Quantum Calabi-Yau and classical crystals,''
arXiv:hep-th/0309208.
} \lref\mss{ A.~Maloney, M.~Spradlin and A.~Strominger,
``Superconformal multi-black hole moduli spaces in four
dimensions,'' JHEP {\bf 0204}, 003 (2002) [arXiv:hep-th/9911001].
}

\lref\BerezinDU{
  F.~A.~Berezin,
  ``General Concept Of Quantization,''
  Commun.\ Math.\ Phys.\  {\bf 40}, 153 (1975).
}

\lref\papad{ G. Papadopoulos, ``Conformal and Superconformal
Mechanics'', hep-th/0002007. }
\lref\rgcv{
R.~Gopakumar and C.~Vafa,
``M-theory and topological strings. I,''
arXiv:hep-th/9809187.
}
\lref\aes{
A.~Strominger,
``Macroscopic Entropy of $N=2$ Extremal Black Holes,''
Phys.\ Lett.\ B {\bf 383}, 39 (1996)
[arXiv:hep-th/9602111].
}
\lref\MohauptMJ{
T.~Mohaupt,
``Black hole entropy, special geometry and strings,''
Fortsch.\ Phys.\  {\bf 49}, 3 (2001)
[arXiv:hep-th/0007195].
}

\lref\VafaGR{
C.~Vafa,
``Black holes and Calabi-Yau threefolds,''
Adv.\ Theor.\ Math.\ Phys.\  {\bf 2}, 207 (1998)
[arXiv:hep-th/9711067].
}
\lref\myers{
R.~C.~Myers,
``Dielectric-branes,''
JHEP {\bf 9912}, 022 (1999)
[arXiv:hep-th/9910053].
}

\lref\myersRV{
R.~C.~Myers,
``Nonabelian Phenomena on D-branes'',
Class.Quant.Grav.20, S347-S372 (2003)
[arXiv:hep-th/0303072].
}

\lref\OoguriZV{ H.~Ooguri, A.~Strominger and C.~Vafa,
``Black hole attractors and the topological string,''
arXiv:hep-th/0405146.
} \lref\spinor{T. Mohaupt, ``Black Hole Entropy, Special Geometry
and Strings", hep-th/0007195.}

\lref\bpsb{ M. Marino, R. Minasian, G. Moore and A. Strominger,
``Nonlinear Instantons from Supersymmetric $p$-Branes",
hep-th/9911206. }

\lref\msw{ J. Maldacena, A. Strominger and E. Witten, ``Black Hole
Entropy in M-Theory", hep-th/9711053. }

\lref\SimonsNM{
A.~Simons, A.~Strominger, D.~M.~Thompson and X.~Yin,
``Supersymmetric branes in AdS(2) x S**2 x CY(3),''
arXiv:hep-th/0406121.
}
\lref\suss{
B.~Freivogel, L.~Susskind and N.~Toumbas,
``A two fluid description of the quantum Hall soliton,''
arXiv:hep-th/0108076.
}

\lref\KontsevichVB{
  M.~Kontsevich,
  ``Deformation quantization of Poisson manifolds, I,''
  Lett.\ Math.\ Phys.\  {\bf 66}, 157 (2003)
  [arXiv:q-alg/9709040].
}

\lref\SpradlinKU{
  M.~Spradlin and A.~Volovich,
  ``Noncommutative solitons on Kaehler manifolds,''
  JHEP {\bf 0203}, 011 (2002)
  [arXiv:hep-th/0106180].
}

\lref\nlsusy{ K. Sugiyama and K. Yoshida,
``Supermembrane on the PP-wave Background'',
Nucl.Phys.B644, 113-127 (2002)
[arXiv:hep-th/0206070]. }

\lref\killing{N. Alonso-Alberca, E. Lozano-Tellechea and T. Ortin,
``Geometric Construction of Killing Spinors and Supersymmetry
Algebra in Homogeneous Spacetimes", hep-th/0208158.}

\lref\poincare{ H. L\"u, C.N. Pope and J. Rahmfeld, ``A Construction
of Killing Spinors on $S^n$'', hep-th/9805151. }

\lref\dzero{ M. Bill\'o, S. Cacciatori, F. Denef, P. Fr\'e,
A. Van Proeyen and D. Zanon, ``The 0-brane action in a general D=4 supergravity
background'', hep-th/9902100. }

\lref\gqkahler{ N. Reshetikhin and L. Takhtajan, ``Deformation
Quantization of K\"ahler Manifolds'', math.QA/9907171. }
\lref\iceland{ R.~Britto-Pacumio, J.~Michelson, A.~Strominger and
A.~Volovich, ``Lectures on superconformal quantum mechanics and
multi-black hole  moduli spaces,'' arXiv:hep-th/9911066.
}

\lref\mss{ A.~Maloney, M.~Spradlin and A.~Strominger,
``Superconformal multi-black hole moduli spaces in four
dimensions,'' JHEP {\bf 0204}, 003 (2002) [arXiv:hep-th/9911001].
}

\lref\mstwo{ J.~Michelson and A.~Strominger, ``Superconformal
multi-black hole quantum mechanics,'' JHEP {\bf 9909}, 005 (1999)
[arXiv:hep-th/9908044].
}

\lref\msone{ J.~Michelson and A.~Strominger, ``The geometry of
(super)conformal quantum mechanics,'' Commun.\ Math.\ Phys.\  {\bf
213}, 1 (2000) [arXiv:hep-th/9907191].
}

\lref\fks{ S.~Ferrara, R.~Kallosh and A.~Strominger, ``N=2
extremal black holes,'' Phys.\ Rev.\ D {\bf 52}, 5412 (1995)
[arXiv:hep-th/9508072].
} \lref\bsv{ R.~Britto-Pacumio, A.~Strominger and A.~Volovich,
``Two-black-hole bound states,'' JHEP {\bf 0103}, 050 (2001)
[arXiv:hep-th/0004017].
}
\lref\bmss{ R.~Britto-Pacumio, A.~Maloney, M.~Stern and
A.~Strominger, ``Spinning bound states of two and three black
holes,'' JHEP {\bf 0111}, 054 (2001) [arXiv:hep-th/0106099].
}

\lref\maldacena{ J. Maldacena, ``The Large N Limit of Superconformal
Field Theories and Supergravity", Adv.Theor.Math.Phys.{\bf 2}, 231
(1998); Int.J.Theor.Phys.{\bf 38}, 1113 (1999)
[arXiv:hep-th/9711200]. }

\lref\kthy{ R. Minasian and G. Moore, ``K-theory
and Ramond-Ramond charge'', hep-th/9710230. }

\lref\gsy{D. Gaiotto, A. Strominger and X. Yin, ``Superconformal
Black Hole Quantum Mechanics'',
  JHEP {\bf 0511}, 017 (2005)
  [arXiv:hep-th/0412322].
}

\lref\deboer{ J. de Boer, A. Pasquinucci and K. Skenderis, ``
AdS/CFT Dualities Involving Large 2D N=4 Superconformal Symmetry'',
Adv.Theor.Math.Phys.{\bf 3}, 577 (1999)
[arXiv: hep-th/9904073]. }


\Title{\vbox{\baselineskip12pt\hbox{hep-th/0412179} }}{ D0-branes
in Black Hole Attractors}

\centerline{Davide Gaiotto, ~Aaron Simons,~ Andrew Strominger and
Xi Yin }
\smallskip
\centerline{Jefferson Physical Laboratory, Harvard University,
Cambridge, MA 02138} \vskip .6in \centerline{\bf Abstract} {
Configurations of $N$ probe D0-branes in a Calabi-Yau black hole
are studied. A large degeneracy of near-horizon bound states are
found which can be described as lowest Landau levels tiling the
horizon of the black hole. These states preserve some of the
enhanced supersymmetry of the near-horizon $AdS_2\times S^2 \times
CY_3$ attractor geometry, but not of the full asymptotically flat
solution. Supersymmetric non-abelian configurations are
constructed which, via the Myers effect,  develop charges
associated with higher-dimensional branes wrapping $CY_3$ cycles.
An $SU(1,1|2)$ superconformal quantum mechanics describing
D0-branes in the attractor geometry is explicitly constructed. }
\vskip .3in

\smallskip
\Date{August 31, 2004}

\listtoc \writetoc

\newsec{Introduction}

Despite a number of exciting chapters, the story of Calabi-Yau
black holes in string theory \foot{An excellent recent review can
be found in \MohauptMJ.} remains to be finished. There are several
indications of this. One is that we have yet to understand the
expected \maldacena\ superconformal quantum mechanics which is
holographically dual to string theory in the near-horizon
$AdS_2\times S^2 \times CY_3$ attractor  geometry. A second is the
mysterious and recently discovered connection \OoguriZV\ relating
the black hole partition function to the square of the topological
string partition function, which in turn is related to the physics
of crystal melting \orv. All of these point to a surprise ending
for the story.

The chapter of interest for this paper involves the study of
D-branes on type II Calabi-Yau black holes with RR charges. This
is relevant for the proposal, pursued in a companion paper \gsy,
that the dual superconformal quantum mechanics may be realized as
D-brane quantum mechanics in the near-horizon attractor geometry.
Branes also play a role in the connection of \OoguriZV\ and are
related to the atoms of the crystals appearing in \orv. Some
development of this chapter appeared recently in \SimonsNM, where
a rich variety of classical, single-brane configurations were
found which preserve some supersymmetry of the near-horizon
attractor region but not of the full black hole geometry.

In this paper we continue this chapter, expanding the classical
analysis of \SimonsNM\ by studying D0-brane configurations,
including quantum effects as well as non-abelian effects for
multiple D0-branes. One interesting result is that, due to
magnetic RR fields, the D0-branes probe a non-commutative
deformation of the attractor geometry. This leads to a large
degeneracy of near horizon D0 bound states, which can be described
as lowest-Landau-level states which tile the black hole
horizon.\foot{Related phenomena were encountered in attempts to
embed the quantum Hall effect into string theory \suss.} This is
suggestive of the old idea that the area-entropy law results from
planckian degrees of freedom which tile the horizon with
planck-sized cells.


This paper is organized as follows. Section 2 describes large $N$
non-abelian collections of D0-branes which grow charges associated
to branes wrapping higher dimensional Calabi-Yau cycles. Parts of
the relevant mathematics - geometric quantization, deformation
quantization and noncommutative geometry - are briefly reviewed in
section 2.1. Section 2.2 embeds these mathematical structures in the
problem of $N$ D0-branes. Supersymmetry (or the equations of motion)
of the D0 configuration gives an interesting constraint on the
noncommutativity parameter $\theta \sim \omega^{-1}$, where $\omega$
is a closed two-form. It is found that, to leading order at large
$N$, the Ricci-flat K\"ahler form $J$ on $CY_3$ must be harmonic
with respect to the metric associated to $\omega$. This result is
shown to agree with that obtained from a DBI analysis of the
D6-brane generated by the D0-brane configuration. In 2.3 formulae
are given for the induced higher-brane charges. In 2.4 an explicit
example at finite $N$ ($N=5$) of a non-abelian D0 configuration on
the quintic is given. In section 3 we turn to D0-branes in the
$AdS_2\times S^2 \times CY_3$ attractor geometry. In 3.1 we show
that the D0 ground states (with respect to the global $AdS_2$ time)
are lowest Landau levels which tile the black hole horizon. The
ground state degeneracy is the D6 flux though the horizon, which we
denote $p_0$.  In 3.2 we construct the bosonic conformal quantum
mechanics describing a D0-brane in $AdS_2\times S^2 \times CY_3$. As
a warm up in 3.3 we supersymmetrize the $AdS_2\times S^2 $ part. In
3.4 the full superconformal quantum mechanics for D0 branes in the
attractor geometry is constructed.  The chiral primary states are
described in 3.5. In 3.6 they are summarized by an index. In 3.7 we
consider the interesting case  a D2 brane with $N$ units of magnetic
flux (i.e. an $N$ D0 brane bound state) wrapping the horizon $S^2$.
The corresponding superconformal mechanics, which plays a role in
the companion paper \gsy, is constructed.

\newsec{$N$ D0-branes in
$R^4\times CY_3$}

In this section we describe configurations of $N$ D0 branes which,
via the Myers effect \myers\ carry charges associated with higher
dimensional branes wrapped around cycles of a Calabi-Yau threefold,
denoted $CY_3$. We will also analyze their supersymmetry properties
in the asymptotic $R^4\times CY_3$ region far from the black hole
where background RR fluxes can be ignored. Near horizon brane
configurations will be considered in the next section.

A collection of $N$ D0-branes with the $N\times N$ matrix
collective coordinates $\Phi^A,~A=1,\cdots 9$ can source higher
dimensional brane charges, through the Chern-Simons terms \refs{\myers,\myersRV}
 \eqn\csa{ \eqalign{ S &= T_0
\int dt {\rm Tr}\left\{ C_t^{(1)} +i{\lambda\over 2}
[\Phi^A,\Phi^B](C_{tAB}^{(3)}+C_t^{(1)}B_{AB}) \right. \cr &
\left.~~~ -{\lambda^2\over 8}[\Phi^A,\Phi^B][\Phi^C,\Phi^D]\left(
C_{tABCD}^{(5)}+C_{t[AB}^{(3)}B_{CD]}+{1\over
2}C_t^{(1)}B_{[AB}B_{CD]} \right)+\cdots\right\} } } where
$\lambda=2\pi\alpha'$. In this expression $C^{(p)}_{\cdots}$ are the various p-form
RR potentials. We will largely consider the case where the NS
potential $B=0$. We are interested in noncommutative D0 brane
configurations in $CY_3$ of the form \eqn\phiconf{
[\Phi^A,\Phi^B]\sim \theta^{AB}(\Phi) ,} where $\theta^{-1} \sim
\omega$ is a non-trivial two-form. In order to give a precise
version of equation \phiconf, it is convenient to employ
 the framework of geometric
quantization, which we now review.

\subsec{Brief review of geometric quantization and the star
product}

Berezin's geometric quantization \refs{\BerezinDU, \SpradlinKU} of a
K\"ahler space $M$ describes the quantum mechanics of a particle
whose phase space is $M$. In this paper we will be interested in the
special case when $M=CY_3$ is a Calabi-Yau threefold. The
quantization begins with a choice of holomorphic line bundle ${\cal
L}$ over $M$, with metric $e^{-K}$ where $K$ is the K\"ahler
potential associated with a chosen K\"ahler form $\omega$. This is
possible only if $[{i\over 2\pi}\omega]=c_1({\cal L})$ is integral.
$\omega$ will turn out to be roughly $\theta^{-1}$. The space of
holomorphic sections of $\cal L$, ${\cal H} = H^0(M,{\cal L})$ is
then a Hilbert space with inner product defined by \eqn\inner{
\langle s_1| s_2\rangle = \int_M {\bar s_1(\bar z)} s_2(z)
e^{-K}d\mu } where $d\mu={1\over 3!}\omega^3$ is the volume form
associated with the K\"ahler form $\omega$. ${\cal H}$ is
finite-dimensional with dimension given by the Riemann-Roch formula
\eqn\dimh{ N\equiv \dim {\cal H} = \chi({\cal O}_M({\cal L})) =
\int_M {1\over 6}\omega^3 + {1\over 12}\omega\wedge c_2(M) }

Let $s_k$ be a basis of holomorphic sections that are orthonormal
with respect to \inner. Then, given an operator ${\cal O}$ acting
on ${\cal H}$, one can get a function (covariant symbol) on $M$
associated with ${\cal O}$ via, \eqn\covsym{ f_{\cal O}(z,\bar z)
= {\sum_{i,j=1}^N \bar s_i(\bar z){\cal O}_{ij}s_j( z) \over
\sum_{i=1}^N |s_i( z)|^2,  } } where \eqn\jig{{\cal
O}_{ij}=\langle s_i|{\cal O}|s_j\rangle} is
 ${\cal O}$ represented as an ${N\times N}$ matrix.
 This gives only $N^2$ independent functions, so not every
 function on $M$ is the symbol of an operator on $\cal H$.
 The space of smooth functions is formally regained from the space of covariant symbols in the limit
 $N\to \infty$.

Instead of working with operators, one can formulate this as
noncommutative geometry on $M$, in which the ordinary algebra of
functions is replaced by the noncommutative $*$ algebra which
reflects the operator algebra on $\cal H$. This is defined for all
functions in the sense of deformation quantization
\refs{\KontsevichVB,\gqkahler}, i.e. in a formal power series
expansion in ${1 \over N}$, or equivalently the noncommutativity
parameter $\theta$ or $\omega^{-1}$. Hence the noncommutative
geometry picture is good for large $N$. To describe this algebra we
first define the Bergman kernel \eqn\bergman{ B(z,\bar z) = \sum_k
s_k(z)\bar s_k(\bar z), } as well as \eqn\ekh{ e(z,\bar z) =
B(z,\bar z) e^{-K(z,\bar z)} } and the Calabi function \eqn\clb{
\phi(z,\bar z;v,\bar v) = K(z,\bar v) +K(v,\bar z) - K(z,\bar z) -
K(v,\bar v) } Note that $e$ and $\phi$ are invariant under the
K\"ahler transformation $K\to K+f+\bar f$. The Calabi function is
well defined in a neighborhood of $M\times \overline M$. The *
product is then given by \eqn\star{ \eqalign{ (f*g)(z,\bar z) & =
\int f(z,\bar v) g(v,\bar z) {B(z,\bar v) B(v,\bar z)\over B(z,\bar
z)} e^{-K(v,\bar v)} d\mu_v \cr &= \int f(z,\bar v) g(v,\bar z)
{e(z,\bar v)e(v,\bar z)\over e(z,\bar z)} e^{\phi(z,\bar z;v,\bar
v)} d\mu_v } } Moreover, the trace of ${\cal O}_f$ is given by
\eqn\trace{{\rm tr} {\cal O}_f = \int f(z,\bar z) B(z,\bar z)
e^{-K}d\mu = \int f(z,\bar z) e(z,\bar z)d\mu } The * algebra
reflects the operator algebra in that \eqn\gax{{\cal O}_{f*g} =
{\cal O}_f{\cal O}_g.}

An differential expression for the $*$ product can be written as
an expansion in ${ 1 \over N}$. To leading order one has
\eqn\stpo{f*g=fg+\theta^{a\bar b}\p_a f \p_{\bar b}
g+\cdots~~~~~~~~~~\theta^{a\bar b}= (\omega^{-1})^{a\bar
b}+\cdots.} The corrections to these expression are given as a
perturbative diagrammatic expansion in \gqkahler. The star
commutators of the complex coordinates themselves obey
\eqn\cfo{[z^a,z^{\bar b}]_*=\theta^{a\bar b}.}

\subsec{Noncommutative configurations at large $N$}

Now we wish to use this structure to describe nontrivial static
configurations of $N$ D0-branes  on $CY_3\times R^4$, in an
expansion in ${1 \over N}$ as in \gqkahler. We consider here only
the leading order behavior. Suppressing for now the $R^4$
coordinates, $N$ D0-brane configurations are described by 6
$N\times N$ matrices which will be denoted $\Phi^A$, $A=1,..6$, .
The matrices $\Phi^A$ can be viewed, with a suitable basis choice,
as operators acting on the Hilbert space $\cal H$ of holomorphic
sections of the line bundle ${\cal L}$ described in the preceding
subsection.\foot{We stress this line bundle $\cal L$ is in general
different from the one associated to the physical Kahler form
$J=ig_{a\bar b}dz^adz^{\bar b}$ on $CY_3$.} As such we can
associate to each matrix $\Phi^A$ the covariant symbol
\eqn\dsa{X^A(z,\bar z) = {\sum_{i,j=1}^N \bar s_i(\bar z)\Phi^A
_{ij}s_j( z) \over \sum_{i=1}^N |s_i( z)|^2. } } According to
\trace\ the matrix trace becomes the integral over $CY_3$, while
from \stpo\ the matrix product of $\Phi^A$ then becomes the star
product of $X^A$ on $CY_3$ \eqn\fxct{ \Phi^A_{ij}\Phi^B_{jk}\to
X^A(z,\bar z)*X^B(z,\bar z).} The star product appearing here is
the one associated to ${\cal L}$ as described above.

We do not expect that \cfo\ is a solution for arbitrary $\theta$, or
equivalently, arbitrary $\omega$. Rather the D0 equation of motion,
or the requirement of unbroken supersymmetry,  should constrain
$\theta$. To find these constraints to leading order for large
$CY_3$, we may expand the D0 action around a point. Locally, the
leading terms are given by the dimensional reduction of d=10 super
Yang-Mills. Supersymmetry requires the familiar $D$ and $F$-flatness
conditions. The reduced action has a superpotential of the form
$W\sim \epsilon_{abc}\Tr \Phi^a\Phi^b\Phi^c $.  The $F$-flatness
condition is the vanishing of  \eqn\fsz{ D_a W  \sim
\epsilon_{abc}[\Phi^b,\Phi^c] , } where $\epsilon$ here is locally
identified as the holomorphic three-form on $CY_3$. $D$-flatness is
the condition \eqn\yyd{g_{a\bar b}\Tr \bigl[ \Phi^a [T_i,\Phi^{\bar
b}] \bigr]=0.} The covariant symbol of \yyd\ is \eqn\rui{g_{a\bar
b}\theta^{a\bar b}=0.} This condition can be rephrased
\eqn\tol{d*_\omega J=0,~~~~J=ig_{a\bar b}dz^adz^{\bar b},} where
$*_\omega$ is the Hodge dual constructed using $\omega$ as a Kahler
form. Hence the physical $J$ is harmonic with respect to $\omega$.
$\omega$ and $\theta$ will be determined in terms of the induced
higher dimensional D-brane charges in the next subsection.

In fact the D-flatness condition \yyd\ is a bit too strong. This
condition arises from demanding the vanishing of the supersymmetry
variation of the worldline fermions, which are in the adjoint of
$U(N)$. This worldline action has four linear and four nonlinear
supersymmetries. The latter act in Goldstone mode as shifts of the
$U(1)$ fermion. \yyd\ arises from demanding an unbroken subgroup
of the linearly realized supersymmetry. However it can also happen
that there is an unbroken supersymmetry which is a linear
combination of the original linear and nonlinear supersymmetries.
This leads to a generalization of \yyd \ with an arbitrary
constant on the right hand side when $T_i$ is the $U(1)$
generator. The covariant symbol of this is the generalization of
\rui \eqn\mmb{g_{a\bar b}\theta^{a\bar b}={\rm constant}.} This
weaker condition still implies \tol, but relaxes the condition
that $\theta$ and $J$ are orthogonal.\foot{ One can also see this directly
from the supersymmetry transformation of the D0-brane world-line fermions \nlsusy,
$\delta \psi = -iD_\tau X^A \gamma_A\epsilon+
\half [X^A, X^B]\gamma_{AB}\epsilon +\eta$, where $\epsilon$
and $\eta$ are covariantly constant spinors on the $CY_3$. }

We will see in the next subsection that this D0 configuration
describes a $CY_3$-wrapped D6 brane with worldvolume magnetic flux
$F\sim\omega \sim \theta^{-1} $. From this point of view we might
have expected $\omega$ to be harmonic which is not what we found
above in \mmb.However the harmonic condition pertains only to
leading order at small $\omega$, while here $\omega$ is large. In
this regime \mmb\ in fact agrees with an analysis of the DBI
action for the D6-brane.
 In the small $\theta$, or
large $F$ limit, the D6-brane world volume BI action takes the
form \eqn\bia{ \eqalign{ S &= \int \sqrt{\det(g+F)} \cr &=\int
\sqrt{\det F} \left[1+{1\over 4}{\rm Tr}(gF^{-1})^2+\cdots\right]
} } When $F$ is of type $(1,1)$, $\int \sqrt{\det F}=\int F^3$ is
topological. We shall also use $F^{-1}$ to denote the 2-form given by
$(F^{-1})^{a\bar b}$ with indices lowered via the metric $g$. \mmb\ implies that
$F^{-1}$ is harmonic with respect to $g$. It follows that $\sqrt{\det F}={
const}\times \sqrt{\det g}$, and ${\rm Tr}(gF^{-1})^2 =\langle
F^{-1}, F^{-1}\rangle$ is also constant. It is then easy to see
that such $F$ satisfies the equation of motion obtained from
varying the second term in the expansion of \bia.

In summary,  to leading order for large $\omega$ and large $CY_3$,
the D0 equations of motion imply that the K\"ahler form $J$ is
harmonic with respect to $\omega$. Hence, \phiconf, for so
constrained $\theta$, is to leading order a static non-abelian
solution for $N$ D0-branes.

\subsec{Induced charges}

For the noncommutative D0-brane configurations \cfo\ the
effective action \csa\ can be rewritten in terms of the covariant
symbols $X^A$ as \eqn\noneff{ \eqalign{ S &= T_0 \int dt
\int_{CY_3} d\mu \,\left\{ C_t^{(1)} +i{\lambda\over 2}
[X^A,X^B]_*(C_{tAB}^{(3)}+C_t^{(1)}B_{AB}) \right. \cr & \left.~~~
-{\lambda^2\over 8}[X^A,X^B]_*[X^C,X^D]_*\left(
C_{tABCD}^{(5)}+C_{t[AB}^{(3)}B_{CD]}+{1\over
2}C_t^{(1)}B_{[AB}B_{CD]} \right) +\cdots\right\}_* } } where we
replaced the trace by the integral over ${CY_3}$ via \trace\ and
 the subscript $*$ means replacing the ordinary products
of the fields by the $*$ product.\foot{We have used $e(z, \bar
z)=1$ for large $N$. Note also that for large $N$,
 $e^{-K}$ will be supported near a few points on ${CY_3}$
corresponding to the position of the $N$ D0-branes. } Ordering
issues will effect the higher order in $\theta$ corrections in
\noneff.

In the small $\theta$ limit, the terms in \noneff\ for which \cfo\ sources Ramond-Ramond
fields may be rewritten
\eqn\rrcoup{ \eqalign{ \int dt\int_{CY_3}
d\mu \, C^{(1)}_t &= N\int dt C_t^{(1)} \cr {1\over 2}\int dt\int_{CY_3}
d\mu \, \theta^{AB}
C^{(3)}_{t AB} &= q_I\int dt
\int_{\alpha^I}C^{(3)}_t
 \cr  {1\over 8}\int  dt\int_{CY_3} d\mu \,
\theta^{AB}\theta^{CD} C^{(5)}_{tABCD} & =
\ p^I\int dt \int_{\beta_I} C^{(5)}_t \cr  {1\over 48} \int  dt\int_{CY_3}
d\mu \, \theta^{AB}\theta^{CD}\theta^{EF}
C^{(7)}_{tABCDEF}& = \int dt \int_{CY_3}
C^{(7)}_t  ,}} where the induced D4 and D2 charges are denoted by
\eqn\conjcharg{ p^I = \int_{\alpha^I}\omega,~~~~ q_I =
\int_{\beta_I}{\omega^2\over 2} } Note that the induced D6-brane
charge is always 1 for this configuration. Hence we may
interpret the D0-brane configuration
$[\Phi^A,\Phi^{B}]=\theta^{AB}$ constructed from the deformation
quantization with K\"ahler form $\omega$ as a single D6-brane
wrapped on ${CY_3}$ with gauge field flux \eqn\jjl{F=\omega.} The
latter has RR charges $Q=e^\omega \sqrt{Td({CY_3})}$ \kthy\ which
agrees with \conjcharg\ at large $N$. The number of D0-branes
contained in a D6-brane with nonzero $F$ is given by the pairing\foot{
This is also the index of the Dirac operator ${\not\!\! D}$ on the Chan-Paton
bundle, which measures the chiral zero modes of open strings
stretched between the D6 and D0-branes.
}
$(Q,\sqrt{Td({CY_3})})=\int_{CY_3} e^\omega \cdot Td({CY_3})$
which is precisely the dimension $N$ in \dimh. In this way the
line bundle ${\cal L}$ we used to construct the Chan-Paton factors
of the D0-branes is naturally interpreted as the $U(1)$ gauge
bundle on the D6-brane. This is a generalization to Calabi-Yau
spaces of the Myers effect \myers\ whereby higher brane charges
are produced from collections of lower dimensional branes.

 \subsec{An example: the quintic}

 In this section we give an explicit example of a non-abelian
 configuration of 5 D0-branes carrying higher brane charges. Consider the quintic 3-fold
$$
X_5 = \{ (z^1)^5+(z^2)^5+(z^3)^5+(z^4)^5+(z^5)^5=0\}\subset {\bf
P}^4
$$
with the K\"ahler form $\omega$ being the curvature of the ${\cal
O}(1)$ bundle. We take $\omega$ as induced from the Fubini-Study
metric on ${\bf P}^4$, i.e. \eqn\fubini{ \omega = {i\over
2\pi}\partial\bar\partial \ln ( \sum_{a=1}^5 |z^a|^2) } We can
write the K\"ahler potential for $\omega$ as $K=\ln
(\sum_a|z^a|^2)$. ${\cal L}={\cal O}_M(1)$ has $N=5$ sections,
corresponding to monomials $z^1,z^2,\cdots,z^5$. The Bergman
kernel is \eqn\bergman{ B(z,\bar z) = c \sum_a z^a\bar z^a } The
$z^a$'s are, of course, defined only up to a local holomorphic
rescaling, i.e. after picking a local trivialization of ${\cal
L}$. The normalization can be determined from \eqn\nordet{ \int B
e^{-K} d\mu = c\cdot {\rm vol}(X_5) = 5 }
Here ${\rm
vol}(X_5)=\int {\omega^3\over 6}=5/6$, so we get $c=6$. The Calabi
function is \eqn\quiclb{ \phi(z,\bar z;v,\bar v) =
\ln{|\sum_{a=1}^5 z^a\bar v^a|^2 \over (\sum_b|z^b|^2)
(\sum_c|v^c|^2)} }
The star product is given by \star\
\eqn\starqui{ (f*g)(z,\bar z) = \int d\mu_v f(z,\bar v)g(v,\bar z)
{6\,|\sum_{a=1}^5 z^a\bar v^a|^2 \over (\sum_b|z^b|^2)
(\sum_c|v^c|^2)} }
Given a suitable function $f(z,\bar z)$, the
associated operator ${\cal O}_f$ acts on a section $s(z)$ as
\eqn\actsec{ \eqalign{({\cal O}_f s)(z) &= \int d\mu_v f(z,\bar v)
B(z,\bar v) s(v)e^{-K(v,\bar v)} \cr &= \int d\mu_v f(z,\bar v)
s(v) {6\sum_a z^a \bar v^a\over \sum_b |v^b|^2} } } In particular,
for functions \eqn\fijfn{ f_{ab}(z,\bar z) = {z^a\bar z^b\over
\sum_{c=1}^5 |z^c|^2} } The associated operators ${\cal O}_{ab}$
acts on section $z^k$ as \eqn\oacts{ {\cal O}_{ab}\cdot z^c = \int
d\mu_v {6z^a\bar v^b v^c\over \sum_d |v^d|^2} = {6\over 5}(\int
d\mu_v) \delta_{bc}z^a = \delta_{bc}z^a } In other words, written
as $5\times 5$ matrices, we have \eqn\omat{ ({\cal O}_{ab})_{mn} =
\delta_{am}\delta_{bn} } In general, a $5\times 5$ matrix $A$
acting on $H^0(X_5;{\cal L})$ is associated with the function
\eqn\afnij{ f_A(z,\bar z) = {\sum_{a,b}A_{ab}z^a \bar z^b\over
\sum_c |z^c|^2} } on $X_5$. These are nothing but the covariant
symbols \covsym.

We can choose a set of local coordinates near say the point
$[z^1,z^2,z^3,z^4,z^5]=[0,0,0,-1,1]$ to be, in an obvious
notation, $f_{15}, f_{25}, f_{35}, f_{51}, f_{52}, f_{53}$. Fixing
$z^5=1$, we define \eqn\ffss{ x^a=f_{a5} = {1\over 2} z^a +{\cal
O}(|z|^3),~~~~  y^a=f_{5a} = {1\over 2}\bar z^a +{\cal
O}(|z|^3),~~~~i=1,2,3. } Their commutation relations are
\eqn\fxxcom{ \eqalign{ & [x^a,x^b]_*=[y^a,y^b]_*=0, \cr & [x^a,y^b]_* =
f_{ab}(x,y)-\delta_{ab}f_{55}(x,y) = -{1\over
2}\delta_{ab} +2x^a y^b+\cdots } } where the omitted terms are
of quadratic or higher order in $x$ and $y$.

Now we may use this to construct a non-abelian configuration of 5
D0-branes which grows higher brane charges. We simply take the
D0-brane matrix $\Phi^i$ ($i=1,\cdots,6$), in the coordinates \ffss, to be the one
whose covariant symbol is $x^a$ or $y^a$, namely :
\eqn\cvb{(\Phi^a)_{mn}=\delta_{am}\delta_{5n},~~~ (\Phi^{a+3})_{mn}
=\delta_{5m}\delta_{an}.}

\cvb\ does not obey the equations of motion \mmb\ for static
solutions. Indeed our construction did not involve the physical
metric $g_{a\bar b}$. Hence the configuration will move under time
evolution. Finding exact static solutions is a more difficult
problem which would require knowing the Ricci-flat metric on the
quintic.

It is straightforward in principle but tedious in practice to
generalize the above construction to $\omega$ being $n$ times the
unit integral K\"ahler form, correspondingly ${\cal L}$ is the
${\cal O}(n)$ bundle. In the case $n\leq 4$, the sections simply
correspond to degree $n$ monomials of the $z^a$'s. When $n\geq 5$,
there are additional relations between the monomials.

\newsec{Superconformal quantum mechanics of D0-branes in $AdS_2\times S^2 \times CY_3$}

Near the horizon of a supersymmetric type II black hole, spacetime
approaches  an $AdS_2\times S^2 \times CY_3$ attractor geometry. In
this region the RR fluxes are important in the D0 dynamics, and the
vector moduli of $CY_3$ approach fixed attractor values which  are
determined by the black hole charges. The attractor equations
governing the moduli $X^\Lambda$ are \fks\aes
\eqn\asf{p^\Lambda={\rm Re}[ C X^\Lambda],~~~~q_\Lambda={\rm
Re}[CF_\Lambda ],} where $\Lambda=0,1 \cdots b_2$, $p_\Lambda$
($q_\Lambda$) are the magnetic (electric) charges of the black hole,
$F_\Lambda$ are the periods of $CY_3$ and $C$ may be set to one by a
gauge transformation.  In this section will consider D0-branes in
these attractor geometries.

 \subsec{Landau levels and the noncommutative horizon}

A single D0-brane couples to the $RR$ gauge field potential via the
first term in \csa\ \eqn\rok{T_0\int dt C^{(1)}_t ,} where $C^{(1)}$
is sourced by D6-branes. In the attractor geometry arising from
$p^0$ D6-branes (as well as other branes), there is a nonzero
two-form magnetic field strength
\eqn\dokj{F^{(2)}=p^0\epsilon_{S2},} where the unit volume form
$\epsilon_{S2}$ integrates to $4\pi$ over $S^2$. This implies, as
for an electron in a magnetic field,  that at low energies the
coordinates of a single D0 become noncommuting
\eqn\ftob{[\Phi^A,\Phi^B]=((F^{(2)})^{-1})^{AB},} and that the low
energy configurations have $S^2$ wave functions described by one of
the $p^0+1$ lowest Landau levels.\foot{ The number of lowest Landau
levels is $p^0+1$ instead of $p^0$ due to the curvature of the
$S^2$. It also follows from the Riemann-Roch formula. } It can be
seen\foot{For example by lifting to the M-theory picture or looking
at the reduced quantum mechanics problem.} that for an appropriate
$AdS_2$ trajectory, as described in \SimonsNM, these lowest Landau
levels preserve one-half, or a total of 4, near-horizon
supersymmetries. Hence a single D0 in an attractor geometry has
$p^0+1$ degenerate supersymmetric ground states
\eqn\dse{\Omega(1)=p^0+1.} These ground states can be pictured as
lowest Landau levels tiling the horizon of the black hole.

  The above result can be generalized to the case of a general probe
brane with four-dimensional magnetic-electric charge vector
$(u^\Lambda,v_\Lambda)$ in an attractor geometry with fluxes
$(p^\Lambda, q_\Lambda)$.  Symplectic invariance fixes the
degeneracy of the lowest Landau level to be \eqn\risp{
e^{S(u,v)}=|p^\Lambda v_\Lambda - q_\Lambda u^\Lambda |.} By
``probe''  brane here we mean that it can be treated as a single
object with no internal moduli in the internal space or
degeneracies (beyond those implied by supersymmetry). This will
not be the case for branes wrapping curves of high degrees in
$CY_3$, which will have additional degeneracies.

\subsec{Conformal quantum mechanics }

We will be interested in a D0-brane moving in the near horizon
$AdS_2\times S^2\times CY_3$ geometry of D0-D4 black hole, with
magnetic-electric charges $(p^0=0, p^A,q_0, q_A=0)$. The bosonic
part of the world-line action takes the form \eqn\wl{ -m\int d\tau +
q\int A  } where $q=m$ as the D0-brane is ``extremal'' in this
background. The radius $R$ of $AdS_2\times S^2$ is equal to the
graviphoton charge $Q$, \eqn\fdtl{ R = Q = (4D q_0)^{1\over 4} l_4 =
{1\over 2}g_s \sqrt{D\alpha'\over q_0} } where $l_4$ is the
four-dimensional Planck length.

The near horizon geometry $AdS_2\times S^2$ of an ${\cal N}=2$ black hole with general
magnetic-electric charges $(p^I, q_I)$
has gauge field strengths $\hat F^I, \hat G_I$ satisfying
\eqn\gaufd{ p^I = {1\over 4\pi}\int_{S^2}{\hat F}^I,~~~q_I = {1\over 4\pi}\int_{S^2}
{\hat G}_I,~~~ \hat G_I^+={\cal N}_{IJ}\hat F^{+J}.
 }
where ${\cal N}_{IJ}$ is the gauge coupling matrix. These relations
can be solved by \eqn\fgex{ \eqalign{ &\hat F^I = p^I \omega_{S^2} +
f^I \omega_{AdS_2},\cr &\hat G_I = q_I \omega_{S^2} + g_I
\omega_{AdS_2}, \cr & CF_I=q_I+i g_I, ~~~CX^I = p^I+i f^I. } }

A superparticle of charges $(u^I, v_I)$ coupled to the gauge fields via
\eqn\ggfc{ \int_{\Sigma} u^I\hat G_I-v_I \hat F^I = q_e R\int {dt\over \sigma}
+ q_m R \int \cos\theta d\phi }
where $\Sigma$ is an auxiliary surface whose boundary is the world line.
We have
\eqn\mnq{\eqalign{ &m = |Z| = |u^IF_I-v_I X^I|,\cr &q_e = {\rm Im} (CZ),\cr
& q_m = {\rm Re}(CZ) = ({p^Iv_I-q_Iu^I}). } }
where we have normalized $|C|=R$.

In the case $q_e=m$, one can rewrite the action in
Poincar\'e coordinate as \eqn\acgl{ -m\int dt \sqrt{ R^2 \left[\sigma^{-2}(
1- \dot{\sigma}^2) -\dot\theta^2- \sin^2\theta \dot\phi^2 \right]- 2g_{a
\bar b}\dot z^a\dot z^{\bar b}  } + mR\int dt {1\over \sigma} } where $g_{a
\bar b}$ is the metric on the CY. Expanding this action to
quadratic order in the time derivatives and replacing $\sigma$
with \eqn\jih{\xi = \sqrt\sigma ,} we can write the action as
\eqn\xact{ m \int dt \left[ 2R\dot\xi^2 + {R\over 2}\xi^2 \left(
\dot\theta^2 + \sin^2\theta\dot\phi^2\right) + {\xi^2\over R} g_{a \bar b}\dot z^a \dot
z^{\bar b}  \right] }

In terms of the conjugate momenta \eqn\conj{\eqalign{ P_\xi
= 4mR\dot \xi, ~~~ P_\theta = mR\xi^2 \dot\theta, ~~~P_\phi = mR\xi^2
\sin^2\theta \dot\phi,~~~ P_{a} = {m\over R} \xi^2 g_{a \bar b}\dot z^{\bar
b} }} we have the Hamiltonian
 \eqn\hamxi{ H = {P_\xi^2 \over 8mR} +
{1\over 2mR\xi^2}\left(P_\theta^2 + {P_\phi^2\over \sin^2\theta}\right)
+ {R\over m\xi^2}P_{a}g^{a \bar b}P_{\bar b}  } A homothety of the full metric
appearing in \hamxi\ is \eqn\hom{ D = \xi{\partial\over
\partial\xi} } The corresponding operator in the conformal quantum
mechanics is given by \eqn\conf{ D={1\over 2}(\xi P_\xi + P_\xi
\xi) } and the special conformal generator is \eqn\spk{ K = 2mR\xi^2
} It is easy to check that $H,D,K$ indeed obey the commutation
relations of $SL(2,{ R})$ \eqn\commsl{ [D,H] = 2iH, ~~~ [D, K] =
-2i K,~~~ [H, K] = -iD }

We can in fact write down the exact Hamiltonian without making the
linear approximation in \acgl. Writing in terms of $\sigma=\xi^2$
\eqn\ffs{ f = 1-\dot\sigma^2 - \sigma^2 (\dot\theta^2 +
\sin^2\theta \dot\phi^2 +{1\over R^2}g_{ij}\dot y^i\dot y^j) } we have
\eqn\cjm{ P_\sigma = mR{\dot\sigma\over \sigma \sqrt{f}},~~~
P_\theta = mR{\sigma\dot\theta\over\sqrt{f}},~~~ P_\phi =
mR{\sigma\sin^2\theta \dot\phi\over\sqrt{f}},~~~ P_i = {m\over R}{\sigma
g_{ij}\dot y^j\over \sqrt{f}}. } The Hamiltonian is then
\eqn\hamex{ H = \sqrt{{m^2R^2\over\sigma^2} + P_\sigma^2 + {1\over
\sigma^2} (P_\theta^2 + {1\over \sin^2\theta}P_\phi^2+ R^2P_i g^{ij}
P_j) } - {mR\over\sigma} } For more general $(q_e, q_m)$,
corresponding to the probe D0-brane having charge not aligned with
the black hole, the RR 1-form potential is \eqn\agen{ A =
q_eR{dt\over \sigma} + q_mR\cos\theta d\phi } Now we have \eqn\pphi{
P_\phi = {mR\sigma \sin^2\theta \dot\phi\over \sqrt{f}}
+q_mR\cos\theta } The full Hamiltonian is then \eqn\hfull{
\eqalign{ H = \sqrt{{m^2R^2\over\sigma^2} + P_\sigma^2 + {1\over
\sigma^2} \left[P_\theta^2 + {1\over
\sin^2\theta}(P_\phi-q_mR\cos\theta)^2+R^2P_i g^{ij} P_j\right] } -
{q_eR\over\sigma} }} The dilatation operator is as usual,
\eqn\ocong{ D = \sigma P_\sigma+P_\sigma \sigma. }

\subsec{Superconformal mechanics on $AdS_2\times S^2$}

As a warm up to the full  $AdS_2\times S^2\times CY_3$ case, in
this subsection we construct the superconformal zerobrane quantum
mechanics of $AdS_2\times S^2 $ without the $CY_3$. This extends
the construction of \mss\msone\ by the addition of a magnetic field
coupling to the zerobrane.

 The collective coordinates of the
zerobrane are the bosonic part of an ${\cal N}=4$  multiplet. In
${\cal N}=1$ language it consists of bosonic superfields ${\bf X}^i$
and fermionic superfield $\Psi$, where \eqn\supa{ {\bf X}^i = x^i -i
\theta \lambda^i,~~~ {\bf \Psi} = i\psi + i\theta b. } A general
${\cal N}=1$ supersymmetric action including coupling to magnetic
field and scalar potential takes the form \papad\ \eqn\actsp{
\eqalign{ S &= \int dt d\theta \left[ {i\over 2}g_{ij}({\bf X})
D{\bf X}^i \dot{\bf X}^j + {1\over 6}c_{ijk}({\bf X}) D{\bf X}^i
D{\bf X}^j D{\bf X}^k \right.\cr & \left.+iA_i({\bf X}) D{\bf X}^i
-{1\over 2}h({\bf X}) \Psi D\Psi + is({\bf X}) \Psi \right] } }
${\cal N}=4$ supersymmetry requires $g_{ij}, c_{ijk}, h$ to be
related to a function $L({\bf X})$ via \mss \eqn\rels{\eqalign{
&g_{ij} =
\partial_i \partial_j L
+{{\epsilon_i}^{mk}}{\epsilon_{jm}}^l\partial_k \partial_l L, \cr
& c_{ijk} = {1\over 2} \epsilon^{pqr}{{\epsilon_p}^l}_i
{{\epsilon_q}^m}_j {{\epsilon_r}^n}_k
\partial_{lmn}L,
\cr & h = \delta^{ij}\partial_i \partial_j L. }} As described in the
previous subsection, the sigma model metric is obtained from the
non-relativistic limit of a particle in the Poincar\'e patch of
$AdS_2\times S^2$, given by \eqn\sigmet{ ds^2 = {dx^k dx^k\over
|\vec x|} } A natural choice of $L$ which produces the metric
\sigmet\ via \rels\ is \eqn\nal{ L(X) = {1\over 2}|\vec X| } It
follows that \eqn\chs{c_{ijk}=0, ~~~h(X)= {1\over |\vec X|}.} The
terms involving the auxiliary field $b$ in \actsp\ is \eqn\bterms{
{1\over 2}h(x) b^2 - s(x)b. } Integrating out $b$ yields the scalar
potential \eqn\scpol{ V(x) = {s(x)^2\over 2h(x)}. } Conformal
invariance requires $V(x)\propto 1/|\vec x|$, so we have \eqn\ssx{
s(X)= {B\over |\vec X|} } for some constant $B$. With \sigmet,\chs,
and \ssx, the action \actsp\ can be written as \eqn\actexpl{ S =
\int dt d\theta \left[ {1\over 2|\vec {\bf X}|}D{\bf X}^i \dot {\bf
X}^i + iA_i({\bf X}) D{\bf X}^i - {1\over 2|\vec {\bf X}|} \Psi
D\Psi + {iB\over |\vec {\bf X}|}\Psi \right]
 }
The 3 additional supercharges act on the ${\cal N}=1$ superfields as
\eqn\exsuv{ \eqalign{ & Q_i {\bf X}^j = {{\epsilon_i}^j}_k D{\bf X}^k + \delta_i^j \Psi,
\cr & Q_i\Psi = i\delta_{ij}\dot {\bf X}^j.  } }
The action \actexpl\ is invariant under $Q_i$'s when $A_i(X)$ is the vector potential
corresponding to magnetic field $F_{ij}=
B\epsilon_{ijk}X^k/|\vec X|^3$, i.e. that of a monopole of charge $B$ at the origin.
We see that ${\cal N}=4$ supersymmetry
determines the potential \scpol\ in terms of the magnetic field
$B$ on the $S^2$. This is expected from the bosonic action one obtains by expanding the
Born-Infeld action near the charged geodesic of the zerobrane, as in section 3.2.


One of the supercharges takes the form \eqn\supch{ \eqalign{ Q &=
-i\lambda^i(\nabla_i + A_i) + \psi s(\vec x) \cr &= -i\lambda^i
(\partial_i - {x^i\over 2|\vec x|^2} + A_i) + \psi {B\over |\vec
x|}   } } where $\nabla$ is the spin connection. The corresponding special
supercharge is \eqn\sssu{ S = \lambda^i D_i =
\lambda^i{x^i\over |\vec x|} }


\subsec{Including the  $CY_3$ factor}
 In this section we give the full $SU(1,1|2)$ quantum mechanics
 for a D0 brane in an attractor geometry.

The metric for D0-brane quantum mechanics on $AdS_2\times S^2\times
CY_3$ is \eqn\mcqm{ \eqalign{ ds^2 &= {4Qd\xi^2}+\xi^2
(Qd\Omega_2^2+{2\over Q}g_{a \bar b}dy^ady^{\bar b}) \cr &= {1\over
|\vec x|}\sum_{k=1}^3dx^k dx^k + {2|\vec x|\over Q^2} g_{a\bar
b}dz^a dz^{\bar b} } } We will work in the coordinate system
$(\xi=\sqrt{\sigma}, \theta,\phi, z^a, z^{\bar a})$. There are four
supercharges $Q_\alpha, \bar Q_\alpha$ and four special supercharges
$S_\alpha, \bar S_\alpha$, where $\alpha=1,2$ is a doublet index
under $SU(2)_R\subset SU(1,1|2)$. The fermions are
$\lambda_\alpha, \bar\lambda_\alpha$ which are roughly the
$\lambda^i, \psi$ appeared in the SCQM for D0-brane on $AdS_2\times
S^2$, and $\eta_\alpha^a, \bar\eta_\alpha^{\bar a}$ which are
roughly the superpartners of $z^a, z^{\bar a}$, $a=1,2,3$. The
$SU(2)$ doublet index will be raised and lowered using
$\epsilon_{\alpha\beta}$, for example, $\lambda^\alpha = \lambda_\beta
\epsilon^{\beta\alpha}$, $\lambda_\alpha =
\epsilon_{\alpha\beta}\lambda^\beta$.
 Let $\hat L_i^{S^2}$
be the angular momentum on the $S^2$. We will be free to trade the
vector index $i$ with a symmetric pair of spinor indices
$(\alpha\beta)$. We shall also define the $SU(2)_R$ generators on
$\lambda$ and $\eta$: \eqn\sutg{\eqalign{ & \hat
L_{\alpha\beta}^\lambda = \lambda_{(\alpha} \bar\lambda_{\beta)},\cr
& \hat L_{\alpha\beta}^\eta = g_{a\bar b}^{CY} \eta_{(\alpha}^a \bar
\eta_{\beta)}^{\bar b}. }} The $SU(2)_{\rm Right}$ generators are
\eqn\rgen{ T_{\alpha\beta} = \hat L_{\alpha\beta}^{S^2} +
L_{\alpha\beta}^\lambda + L_{\alpha\beta}^\eta } The supercharges
are then given by \eqn\superchs{ \eqalign{ & Q_\alpha =
Q^{-\half}\left[{1\over 2}\lambda_\alpha \hat P_\xi -{i\over \xi}
(L_{\alpha\beta}^{S^2} + L_{\alpha\beta}^\eta)\lambda^\beta +
{i\over 4\xi}\bar\lambda_\alpha \lambda^2
+{i\over4\xi}\lambda_\alpha \right] + {\sqrt{2Q}\over
\xi}\eta_\alpha^a \hat P_a^{CY},
 \cr
& \bar Q_\alpha = Q^{-\half}\left[{1\over 2}\bar\lambda_\alpha \hat
P_\xi -{i\over \xi} (L_{\alpha\beta}^{S^2} + L_{\alpha\beta}^\eta)
\bar\lambda^\beta -{i\over 4\xi} \bar\lambda^2 \lambda_\alpha -
{i\over 4\xi}\bar\lambda_\alpha \right] + {\sqrt{2Q}\over \xi}\bar
\eta_\alpha^{\bar a} \hat P_{\bar a}^{CY},\cr & S_\alpha =
2Q^\half\xi \lambda_\alpha,~~~~\bar S_\alpha =
2Q^\half\xi\bar\lambda_\alpha. } } where $\lambda^2\equiv
\lambda_\beta\lambda^\beta$, $\bar\lambda^2 \equiv \bar\lambda_\beta
\bar\lambda^\beta$. They satisfy commutation relations
\eqn\commsucf{ \eqalign{ &\{ Q_\alpha, Q_\beta\} = 0,~~~~~
\{Q_\alpha, \bar Q_\beta\} = 2\epsilon_{\alpha\beta} H,\cr
&\{S_\alpha, \bar S_\beta\}=2\epsilon_{\alpha\beta}K,~~~~~
\{Q_\alpha, \bar S_\beta\} = \epsilon_{\alpha\beta}D
-2iT_{\alpha\beta}. } }  The wave functions on the CY can be
identified with forms as usual. The doublets $\eta^a_\alpha
P_a,\bar\eta^{\bar a}_\alpha P_{\bar a}$ are then identified with
differential operators on the CY as \eqn\suld{ \left( \eqalign{
\eta^a_1 P_a \cr \eta^a_2 P_a } \right) \to \left( \eqalign{
&\partial \cr &\bar\partial^* } \right),~~~~ \left( \eqalign{
\bar\eta^{\bar a}_1 P_{\bar a} \cr \bar\eta^{\bar a}_2 P_{\bar a} }
\right)\to \left( \eqalign{\bar\partial~~& \cr -\partial^*& }
\right) } $\hat L^\eta_{\alpha\beta}$ are identified with the
generators of Lefschetz action on the forms. Note that in \superchs\
the angular momentum of the D0-brane on the $S^2$ is curiously
twisted by the Lefschetz weight of the state. In addition to
$SU(2)_R$, we have an additional $SU(2)_L$ R-symmetry which rotates
$Q,\bar Q$ as a doublet. The $SU(2)_L$ generators are given by
\eqn\sutlgen{ \eqalign{ &J_L^+ = \lambda^2 + P_{{\cal H}^\perp}
{1\over\Delta}\bar\partial\partial^* ,\cr &J_L^- = \bar\lambda^2 +
P_{{\cal H}^\perp} {1\over\Delta}\partial\bar\partial^*,\cr & J_L^3
= {1\over 2}\lambda\bar\lambda-{1\over 2} + P_{{\cal
H}^\perp}{1\over 2\Delta}(
\partial^*\partial-\bar\partial^*\bar\partial).
} } where $\Delta$ is the Laplacian, $P_{{\cal H}^\perp}$ is the
projection onto the orthogonal of the harmonic space. Note that
the RHS of \sutlgen\ is well defined since $\Delta^{-1}$ is
bounded on ${\cal H}^\perp$ for compact $CY_3$. Together with the
$SU(1,1|2)$ constructed above we have the full $D(2,1;0)$
superconformal algebra.

In the $SU(2)_R\times SU(2)_L$ doublet notation, we shall write
\eqn\pmdb{ Q_1 = Q^{++}, ~~~Q_2 = Q^{-+},~~~ \bar Q_1 = Q^{+-},~~~
\bar Q_2 = Q^{--}. } Now \eqn\qpmas{ \eqalign{ &Q^{\pm +} =
Q^{-\half}\left[{1\over 2}\lambda^{\pm +} \hat P_\xi - {i\over \xi}
({L^{S^2\pm}}_{\alpha} +{L^{\eta\pm}}_\alpha) \lambda^{\alpha +} +
{i\over 4\xi}\{\lambda^{\pm -},\lambda^{-+}\lambda^{++} \} \right]
+{\sqrt{2Q}\over \xi}\eta^{\pm +}\cdot \hat P^{CY}, \cr
 &Q^{\pm -}
= Q^{-\half}\left[{1\over 2}\lambda^{\pm -} \hat P_\xi - {i\over
\xi} ({L^{S^2\pm}}_{\alpha} +{L^{\eta\pm}}_\alpha) \lambda^{\alpha
-} - {i\over 4\xi}\{\lambda^{\pm +},\lambda^{--}\lambda^{+-} \}
\right] +{\sqrt{2Q}\over \xi}\eta^{\pm -}\cdot \hat P^{CY}, \cr
&S^{\pm\pm}=2Q^\half\xi\lambda^{\pm\pm}. } }

\subsec{Chiral primaries}

In this section we describe the chiral primaries states of the
superconformal quantum mechanics.  Related discussions appear in
\refs{\bsv, \bmss}.

Let us define \eqn\gchs{ G_{\pm \half}^{\alpha A} =
{1\over\sqrt{2}}\left(Q^{\alpha A} \mp i S^{\alpha A}\right). }
where $\alpha, A=\pm$. We shall look for chiral primaries, which are
states annihilated by the supercharges \eqn\bpchs{
G_{\pm\half}^{++},~~~~G_{\pm\half}^{+-}, ~~~~ G_{\half}^{-+}, ~~~~
G_{\half}^{--}.  } Such a state $|0\rangle$ must be annihilated by
$S^{+A}$, hence $\lambda^{+A}$, for $A = \pm$. We can demand that
the $CY_3$ part of $|0\rangle$ is a harmonic form, i.e. it is
annihilated by $\eta^{\pm\pm}\cdot\hat P^{CY}$. With
$T^{++}|0\rangle=0$, i.e. $|0\rangle$ has the highest $SU(2)_R$
weight, the condition $G_{\pm \half}^{+A}|0\rangle = 0$ is
automatically satisfied. Requiring $G_{\half}^{-\pm}|0\rangle = 0$
now gives \eqn\diffeq{ \eqalign{ & (-{i\over 2}\lambda^{-+}
\partial_\xi + {i\over \xi}(L^{S^2}+L^\eta)^{-+}\lambda^{-+}
+{i\over 4\xi}\lambda^{-+}-2iQ\xi\lambda^{-+} )|0\rangle = 0, \cr &
(-{i\over 2}\lambda^{--}\partial_\xi +{i\over
\xi}(L^{S^2}+L^\eta)^{-+}\lambda^{--} +{i\over
4\xi}\lambda^{--}-2iQ\xi\lambda^{--} )|0\rangle = 0. } } The third
$SU(2)_R$ generator is $J_R^3 = T^{-+}= (L^{S^2}+L^\eta)^{-+}+\half$
when acting on $|0\rangle$. We conclude that a general chiral
primary state takes the form \eqn\wvfngr{ |0\rangle\sim
e^{-2Q\xi^2}\xi^{2j_R-\half} |j_R, J_R^3=j_R; L^{S^2},
L^\lambda=\half, \omega_{CY}\rangle
 }
where $\omega_{CY}$ is a primitive harmonic form on the $CY_3$. For example,
if we identify $(L^\eta)^{++}=i_J$ where $J$ is the K\"ahler form, the primitive harmonic
forms are 1 (in the spin $3/2$ representation), $h^{11}-1$ harmonic 2-forms $\omega$
satisfying $\langle J, \omega\rangle = 0$ (in the spin $1/2$ representation),
and $2(h^{21}+1)$ harmonic 3-forms (spin 0).

Now let us add magnetic field $F_{ij} = B\epsilon_{ijk}x^k/|x|^3$. If we replace the standard
derivatives by gauge covariant derivatives, the angular momenta $L^{S^2}_i$
gets modified to $\tilde L^B_i$, which satisfy commutation relations
\eqn\blstt{ [\tilde L_i^B, \tilde L_j^B] = i\epsilon_{ijk}\left(\tilde L_k^B + B{x^k\over|x|}\right) }
We can further define
\eqn\ltill{ L^B_k = \tilde L^B_k - {Bx^k\over 2|x|} }
so that $L_k^B$ now satisfy the standard $SU(2)$ algebra.
The supercharges are given by the expression \superchs\ with $L^{S^2}$ replaced
by $L^B$. Note that $(L^B)^2=(\tilde L^B)^2+B^2/4$.
The chiral primaries are still given by \wvfngr, with $L^{S^2}$ replaced by $L^B$. The only
difference is that we must restrict the spin $L^B= {B\over 2}, {B\over 2}+1, {B\over 2}+2,\cdots$.

We have another $U(1)$ R-symmetry that rotates the
$\lambda,\eta$'s and $\bar \lambda,\bar\eta$'s with opposite phases. This $U(1)$ generator is
\eqn\uogen{ \tilde J = {1\over 2}\lambda_\alpha\bar\lambda^\alpha + {1\over 2}g_{a\bar b}\eta^a_\alpha
 \bar\eta^{\bar b\alpha} - 1 } The chiral primaries states \wvfngr\
have $\tilde J = (p-q)/2$ if $\omega_{CY}$ is a $(p,q)$-form. $\tilde J$ doesn't commute with
$J_L^\pm$. We can however take the linear combination $J_0\equiv \tilde J-J_L^3$, which commutes
with $SU(2)_L$. The corresponding $U(1)$ symmetry will be denoted $U(1)_0$.

We can assemble the chiral primaries into representations of
$SU(2)_L\times SU(2)_R\times U(1)_0$, where $SU(2)_R$ is the
R-symmetry appearing in $SU(1,1|2)$. The spectrum of the chiral
primaries is \eqn\chiralp{ \eqalign{ &\bigoplus_{n=0}^\infty \left[
\left( 0, {B\over 2}+n+{1\over 2}, {3\over 2} \right) \oplus
h^{21}\left( 0, {B\over 2}+n+{1\over 2}, {1\over 2} \right)
\right.\cr &\left.~~~ \oplus h^{21} \left( 0, {B\over 2}+n+{1\over
2}, -{1\over 2} \right) \oplus \left( 0, {B\over 2}+n+{1\over 2},
-{3\over 2} \right)\right. \cr &\left.~~~\oplus \left(0, ({B\over
2}+n)\otimes ({3\over 2})+{1\over 2}, 0\right) \oplus (h^{11}-1)
\left( 0, ({B\over 2}+n)\otimes ({1\over 2})+{1\over 2}, 0 \right)
\right] } } where we used the shorthand notation $(n)\otimes (m)$
for all the $SU(2)$ highest weights appearing in the tensor product
$(n)\otimes (m)$. In the case $B=0$, we can write the chiral
primaries labeled by $SU(2)_L\times SU(2)_R$ representations as
\eqn\chiralsimp{ \bigoplus_{n=0}^\infty  \left[ h^{odd}(0, n+\half)
\oplus h^{even}(0, n+1) \right] }


The chiral primaries sit in short representations of $D(2,1;0)$,
denoted by $(j_L,j_R)_S$, which has spin contents\deboer\ (see also appendix
A)
\eqn\shortreps{ \eqalign{ (j_L,j_R)_S \to (j_L,j_R)_0
+(j_L+\half,j_R-\half)_{{1\over 2}} + (j_L-\half,j_R-\half)_{{1\over
2}} + (j_L,j_R-1)_{1} } } where
the subscripts label the difference of the $L_0$ value of the
corresponding $SU(2)\times SU(2)$ representation from that of the
chiral primary. The chiral primaries in our SCQM have $j_L=0$, so the relevant
short representation spin content is
\eqn\shortzero{ \eqalign{
(0,j_R)_S \to (0,j_R)_0+(\half, j_R-\half)_{1\over 2} + (0,j_R-1)_1
} } States in a short multiplet are obtained by acting
$G_{-1/2}^{\pm\pm}$ on the chiral primary. We have $L_0 = j_R$ for
the chiral primary state, and with each action of $G_{-1/2}$, $L_0$
increases by $1/2$.

\subsec{An index} Information about the chiral primaries can be
summarized by an index which we now discuss.

Other, non-chiral-primary, states in the superconformal quantum
mechanics form long representations $(j_L,j_R)_{long}$, which has
the spin content of four short multiplets \eqn\longrep{
(j_L,j_R)_{long} \to (j_L,j_R)_S + (j_L-\half,j_R+\half)_S
+ (j_L+\half,j_R+\half)_S + (j_L,j_R+1)_S } A
general long multiplet satisfies the BPS bound $L_0\geq j_R$. A
generic generating function ${\rm Tr}y^{J_L^3}z^{J_R^3}q^{L_0}$
receives contribution from both short and long multiplets. We can,
however, construct the index ${\rm Tr} (-1)^{2J_L^3}
y^{L_0+J_L^3}z^{L_0+J_R^3}$ which vanishes for long multiplets but
doesn't vanish for short multiplets. Evaluating the trace over short
multiplets gives \eqn\shorindx{ {\rm Tr}_{(j_L,j_R)_S}
(-1)^{2J_L^3}y^{L_0+J_L^3}z^{L_0+J_R^3} =
(-1)^{2j_L}y^{j_R-j_L}{(1-yz)(1-y^{2j_L+1})\over 1-y} } We are free to insert $w^{J_0}$ into the
trace to get a slightly refined index. In the case $B=0$, it is
straightforward to calculate from \chiralp\ \eqn\genind{ \eqalign{
{\rm Tr} (-1)^{2J_L^3+2J_0}y^{L_0+J_L^3} z^{L_0+J_R^3} w^{J_0} &=
{1-yz\over 1-y}\sum_{r,s=0}^3 (-1)^{r+s} h^{r,s}
y^{\delta_{r,s}+\half} w^{r-s\over 2} } } where \eqn\delrs{
\delta_{r,s} \equiv \left\{ \eqalign{& 0,~~~~ r+s={\rm odd},\cr
&{1\over 2},~~~~ r+s={\rm even}. } \right. } There is no essential
difficulty in generalizing to the case $B\not=0$, but the formula
for the D0-brane index would be messier. So we will stay with $B=0$
in the discussions below.


We can define an index that includes all multiple D0-brane states,
\eqn\fullindes{ {\rm Tr} (-1)^{2J_L^3+2J_0} y^{L_0+J_L^3}
z^{L_0+J_R^3} w^{J_0} q^{N} } where $N$ is the total D0-brane
charge. For a bound state of $k$ D0-branes, the only change in the
index is that the magnetic field $B$ is effectively replaced by
$kB$.

The full index of multi-D0-branes in $AdS_2\times S^2\times CY_3$ is
\eqn\fullind{ \eqalign{ &{\rm Tr} (-1)^{2J_L^3+2J_0}y^{L_0+J_L^3}
z^{L_0+J_R^3} w^{J_0} q^{N} \cr &= \prod_{k=1}^\infty
\prod_{r,s=0}^3 \prod_{n=0}^\infty \left({ 1-q^k
y^{n+\delta_{r,s}+{3\over 2}}z w^{r-s\over 2}\over 1-q^k
y^{n+\delta_{r,s}+{1\over 2}}w^{r-s\over 2}
}\right)^{(-1)^{r+s}h^{r,s}} } } where we traced over all possible
numbers of various D0-brane bound states. In deriving the above
expression we used the fact that the fermion number $(-1)^{ F}$ of
the chiral primaries is essentially $(-1)^{r+s}$, and for the
descendants $(-1)^{ F}$ changes according to $(-1)^{2J_L^3}$.

If we  set $w=1$ and $z=1$, the index \fullind\ can be written
\eqn\indsim{ \eqalign{ &{\rm Tr} (-1)^{2J_L^3+2J_0}y^{L_0+J_L^3}
q^{N}\cr &= \prod_{k=1}^\infty  \left({ 1-q^k y}\right)^{-h_{even}}
\left({ 1- q^k y^{\half}}\right)^{h_{odd}} } } An interesting
special case is $y=e^{2\pi i}$. \indsim\ then becomes \eqn\risx{\Tr
(-)^{F+2J_0}q^N= \prod_{k=1}^\infty \left({ 1- q^k
}\right)^{-h_{even}} \left({ 1+ q^k }\right)^{h_{odd}}.} In fact
\risx\ is true for general magnetic flux $B$ on the $S^2$, if we
replace $q$ by $\tilde q = (-1)^Bq$.

\subsec{Wrapping the horizon}

In this section we construct the superconformal quantum mechanics
for  a non-abelian configuration of $N$ D0-branes which via the
Myers effect becomes a D2 brane with $N$ units of worldvolume
magnetic flux wrapping the horizon.  These configurations play an
important role in black hole entropy as described in \gsy.

  The low energy limit of the
world-volume theory of the D2-brane expanded near its geodesic can
be described again by a ${\cal N}=4$ superconformal quantum
mechanics. Since it wraps the $S^2$, the D2-brane sees an
$AdS_2\times CY_3$ target space. There are no world-volume zero
modes of the gauge field. Due to the background  RR 4-form flux of
the form $F^{(4)}=\omega_{S^2}\wedge J$, where $J$ is the K\"ahler
form on $CY_3$, the $S^2$-wrapped D2-brane  is effectively charged
under a magnetic field $F_{CY}=4\pi QJ$ on the CY (here $Q$ is the graviphoton
charge, not to be confused with supercharges). The
supercharges are of the form \eqn\sschdt{ \eqalign{ & Q_\alpha =
(QT_{Q,N})^{-\half}\left[{1\over 2}\lambda_\alpha \hat P_\xi
-{i\over \xi} L_{\alpha\beta}^\eta\lambda^\beta + {i\over
4\xi}\bar\lambda_\alpha \lambda^2
+{i\over4\xi}\lambda_\alpha\right]\cr &~~~~~~ + \left( {Q\over
T_{Q,N}}\right)^\half \left[{\sqrt{2}\over \xi}\eta_\alpha^a (\hat
P_a^{CY}+A_a) - 4\pi Q^2 {i\over\xi}\lambda_\alpha \right],
 \cr
& \bar Q_\alpha = (QT_{Q,N})^{-\half}\left[{1\over
2}\bar\lambda_\alpha \hat P_\xi -{i\over \xi} L_{\alpha\beta}^\eta
\bar\lambda^\beta -{i\over 4\xi} \bar\lambda^2 \lambda_\alpha -
{i\over 4\xi}\bar\lambda_\alpha \right] \cr &~~~~~~ + \left( {Q\over
T_{Q,N}}\right)^\half \left[ {\sqrt{2}\over \xi}\bar
\eta_\alpha^{\bar a} (\hat P_{\bar a}^{CY}+A_{\bar a})+ 4\pi Q^2
{i\over\xi}\bar \lambda_\alpha \right],\cr & S_\alpha =
(QT_{Q,N})^{\half} 2\xi \lambda_\alpha,~~~~\bar S_\alpha =
(QT_{Q,N})^{\half}2\xi\bar\lambda_\alpha. } } where $A$ is the gauge
connection on the CY, with field strength $dA=F_{CY}=4\pi QJ$.
$T_{Q,N}=\sqrt{(4\pi Q^2)^2+N^2}$ is the mass of the D2-brane with
$N$ D0-brane charges. Comparing with \superchs, we dropped the
$L^{S^2}$ terms, replaced the $\hat P^{CY}$ by the gauge covariant
derivative, and added $-iK^{-1} S_\alpha = -i\xi^{-1}\lambda_\alpha$
to $Q_\alpha$ and $iK^{-1} \bar S_\alpha = i\xi^{-1}\bar
\lambda_\alpha$ to $\bar Q_\alpha$. The last modification is crucial
for $\{ Q_\alpha,\bar Q_\beta\} = 2\epsilon_{\alpha\beta} H$ to hold
for some Hamiltonian $H$. Most of the superalgebra is the same as
$D(2,1;0)$ given by \commsucf, except for \eqn\commnew{ \eqalign{
&\{ Q_\alpha, \bar S_\beta\} = \epsilon_{\alpha\beta}(D-8\pi iQ^3) -
2i T_{\alpha\beta},\cr & \{ S_\alpha, \bar Q_\beta\} =
\epsilon_{\alpha\beta}(D+8\pi iQ^3) + 2i T_{\alpha\beta}. } } where
$T_{\alpha\beta} = L^\eta_{\alpha\beta}+L^\lambda_{\alpha\beta}$.
Writing the supercharges as $SU(2)_R\times SU(2)_L$ doublets,
$Q_{\alpha+}\equiv Q_\alpha, Q_{\alpha-}\equiv \bar Q_\alpha$, etc.,
we have \eqn\commdbn{ \{ Q_{\alpha A}, S_{\beta B}\} =
\epsilon_{\alpha\beta} \epsilon_{AB} D -2i \epsilon_{AB}
T_{\alpha\beta} - i\epsilon_{\alpha\beta} R_{AB} } where
$R_{+-}=R_{-+}=8\pi Q^3, R_{++}=R_{--}=0$. The commutators of
$G_{\pm \half}^{\alpha A}\equiv {1\over \sqrt{2}}(Q^{\alpha A}\mp
iS^{\alpha A})$ are \eqn\comggs{ \eqalign{ & \{ G_{\pm
\half}^{\alpha A}, G_{\pm \half}^{\beta B} \} =
\epsilon^{\alpha\beta} \epsilon^{AB} (H-K\mp iD), \cr &\{
G_{\half}^{\alpha A}, G_{-\half}^{\beta B} \} =
\epsilon^{\alpha\beta} \epsilon^{AB} (H+K) + 2\epsilon^{AB}
T^{\alpha\beta} +\epsilon^{\alpha\beta} R^{AB}. } } This is
$SU(1,1|2)$ algebra with a central extension $R_{AB}$, which
explicitly breaks $SU(2)_L$ R-symmetry.

An interesting problem is to compute an index that counts the chiral
primaries of this superconformal quantum mechanics, analogous to the
one in section 3.6. It is shown in a companion paper \gsy\ that
these chiral primaries have the right degeneracies to account for
the leading entropy of D4-D0 black holes, suggesting that the
horizon wrapped D2-branes may describe the miscrostates of the black
hole.

\centerline{\bf Acknowledgements} This work was supported in part by
DOE grant DE-FG02-91ER40654. We are grateful to D. Jafferis, D.
Thompson and C. Vafa for useful conversations.

\appendix{A}{Representations of ${\cal N}=4$ superconformal algebras}

In this appendix we describe some properties of representations of the superconformal
algebras related to D0-branes in $AdS_2\times S^2$.

\subsec{$D(2,1;\alpha)$ with $\alpha\not=0$}
The superconformal algebra $D(2,1;\alpha)$ takes the form
\eqn\dtosca{
\eqalign{
&\{ G_{\pm \half}^{\alpha A}, G_{\pm \half}^{\beta B} \}
= \epsilon^{\alpha\beta}\epsilon^{AB} L_{\pm 1},
\cr
& \{ G_{\half}^{\alpha A}, G_{-\half}^{\beta B} \}
= \epsilon^{\alpha\beta} \epsilon^{AB} L_0 +
\gamma\epsilon^{\alpha\beta}T_L^{AB}+(1-\gamma)\epsilon^{AB}T_R^{\alpha\beta}.
}
}
where $\gamma=\alpha/(1+\alpha)$. Note that $(G_{\pm \half}^{\alpha A})^\dagger =
\epsilon_{\alpha\beta}\epsilon_{AB}G_{\mp\half}^{\beta B}$.
A highest weight state $|h\rangle$ is annihilated by all the $G_{\half}^{\alpha A}$'s.
$|h\rangle$ is a chiral primary if it is further annihilated by $G_{-\half}^{++}$, or
equivalently if $L_0|h\rangle = (\gamma j_L + (1-\gamma) j_R) |h\rangle$.
Chiral primaries are contained in short representations, with $SU(2)_L\times SU(2)_R$ spin content\deboer
\eqn\dtoshort{ \eqalign{ (j_L,j_R)_S \to (j_L,j_R)_0
+(j_L+\half,j_R-\half)_{{1\over 2}} + (j_L-\half,j_R-\half)_{{1\over
2}}  + (j_L-\half,j_R+\half)_{{1\over 2}} \cr  +
(j_L-1,j_R)_{1}+(j_L,j_R-1)_{1}+ (j_L, j_R)_{1}+
(j_L-\half,j_R-\half)_{{3\over 2}} } }
where the subscripts denote the increase of $L_0$ value compared to the chiral primary $|h\rangle$,
which is the highest weight state in $(j_L,j_R)_0$. The eight highest $SU(2)\times SU(2)$ states in \dtoshort\
are obtained from $|h\rangle$ by acting with the three broken $G_{-1/2}$'s. In writing \dtoshort\ we omitted
$SL(2,{\bf R})$ descendants of $|h\rangle$ and the states obtained by acting on them with
$G_{-\half}$'s. These descendants have the spin content similar to a long representation.
A general long representation $(j_L,j_R)_{long}$ satisfy the unitarity bound
$L_0\geq \gamma j_L + (1-\gamma) j_R$, and has the spin content of short representations
$(j_L,j_R)_S + (j_L+\half, j_R+\half)_S$.

A general index that vanishes for long representations and keeps informations about
short representations of $D(2,1;\alpha)$ is of the form
\eqn\genloind{
{\rm Tr} (-1)^{2J_L^3} y^{L_0\pm J_L^3} z^{L_0\pm J_R^3}
}
This will be an index for the representations of superconformal algebras discussed below
as well.

\subsec{$D(2,1;0)$}
A highest weight state $|h\rangle$ of $D(2,1;0)$ that is annihilated by $G_{-\half}^{++}$
is necessarily annihilated by $G_{-\half}^{+-}$. These are the chiral primaries of
$D(2,1;0)$, which generate short representations with spin content
\eqn\dtozshort{ \eqalign{ (j_L,j_R)_S \to (j_L,j_R)_0
+(j_L-\half,j_R-\half)_{{1\over 2}} + (j_L+\half,j_R-\half)_{{1\over
2}}  +  (j_L,j_R-1)_{1} } } A long representation has spin content of
four short representations
\eqn\dzzlong{
(j_L,j_R)_{long} \to (j_L,j_R)_S + (j_L-\half, j_R+\half)_S +
(j_L+\half, j_R+\half)_S + (j_L, j_R+1)_S
}

\subsec{$SU(1,1|2)$}
$D(2,1;0)$ is $SU(1,1|2)$ together with its outer automorphism $SU(2)_L$. The R-symmetry
of $SU(1,1|2)$ is denoted $SU(2)_R$. Similar to $D(2,1;0)$, a short representation of
$SU(1,1|2)$ has $U(1)_L\times SU(2)_R$ spin content
\eqn\sushort{ \eqalign{ (j_L^3,j_R)_S \to (j_L^3,j_R)_0
+(j_L^3-\half,j_R-\half)_{{1\over 2}} + (j_L^3+\half,j_R-\half)_{{1\over
2}}  +  (j_L^3,j_R-1)_{1} } } where we labelled the $J_L^3$ charge of $SU(2)_L$ for
each $SU(2)_R$ multiplet. A long representation again consists of four short
representations.

\subsec{$SU(1,1|2)$ with central extension}
As discussed in section 3.7, the D2-brane wrapped on the $S^2$ has a superconformal algebra which is a central extension
of $SU(1,1|2)$. We shall denote it by $SU(1,1|2)_Z$. The anti-commutators are given by
\eqn\suzsca{
\eqalign{
&\{ G_{\pm \half}^{\alpha A}, G_{\pm \half}^{\beta B} \}
= \epsilon^{\alpha\beta}\epsilon^{AB} L_{\pm 1},
\cr
& \{ G_{\half}^{\alpha A}, G_{-\half}^{\beta B} \}
= \epsilon^{\alpha\beta} \epsilon^{AB} L_0 +
\epsilon^{AB}T^{\alpha\beta} + \epsilon^{\alpha\beta} R^{AB}.
}
}
where $R^{AB}$ is a constant symmetric tensor, with $R^{++}=R^{--}=0$,
$R^{+-}=R^{-+}=r>0$. Now a chiral primary state $|h\rangle$ can only be defined to be
annihilated by $G_{-\half}^{++}$ but not by $G_{-\half}^{+-}$. It
saturates the unitarity bound $L_0|h\rangle = (j_R+r)|h\rangle $. The short representations
have spin content similar to \dtoshort, except that
we can only label them by $U(1)_L\times SU(2)_R$ multiplets,
\eqn\suzshort{ \eqalign{ (j_L^3,j_R)_S \to (j_L^3,j_R)_0
+(j_L^3+\half,j_R-\half)_{{1\over 2}} + (j_L^3-\half,j_R-\half)_{{1\over
2}}  + (j_L^3-\half,j_R+\half)_{{1\over 2}} \cr  +
(j_L^3-1,j_R)_{1}+(j_L^3,j_R-1)_{1}+ (j_L^3, j_R)_{1}+
(j_L^3-\half,j_R-\half)_{{3\over 2}} } }
Note that the highest weight state in $(j_L^3-\half, j_R+\half)_\half$ also saturates the unitarity bound,
but it is not annihilated by $G_\half^{-+}$. A long representation
$(j_L^3, j_R)_{long}$ of $SU(1,1|2)_Z$ has the spin content of two short representations
$(j_L^3,j_R)_S + (j_L^3+\half,j_R+\half)_S$.

\listrefs

\end